\documentclass{ws-procs9x6}

\begin{document}

\title{X-ray Evidence for Multiple Absorbing Structures in Seyfert
  Galaxies\footnote{\uppercase{D}rawn from
  \uppercase{C}hap.~4 of \uppercase{G}elbord
  2002\protect\cite{gelbord02}, wherein more details and references
  may be found.  \uppercase{O}nline at
  http://space.mit.edu/$\sim$jonathan/papers/thesis/abstract.html.}}

\author{J.~M. GELBORD}

\address{MIT Center for Space Research, NE80-6091\\
77 Massachusetts Ave. \\ 
Cambridge, MA 02139, USA\\ 
E-mail: jonathan@space.mit.edu}

\author{K.~A. WEAVER and T. YAQOOB}

% \address{World Scientific Publishing Co Ltd, \\ 
% 57 Shelton Street, \\
% London WC2H 9HE, England\\
% E-mail: wspc@wspc.ox.uk}  


\maketitle

\abstracts{
We have used X-ray spectra to measure attenuating columns in a large
sample of Seyfert galaxies.  Over 30 of these sources have resolved
radio jets, allowing the relative orientation of the nucleus and host
galaxy to be constrained.  We have discovered that the distribution of
absorbing columns is strongly correlated with the relative orientation
of the Seyfert structures.  This result is inconsistent with
unification models having only a torus and is instead most readily
% the canonical unified model of Seyferts and is instead most readily
explained if a second absorber is included: in addition
to a Compton-thick
% parsec-scale
torus there would also be a
larger-scale absorber with $N_H < 10^{23}$~cm$^{-2}$.
The second absorber is aligned with the
host galactic plane while the torus is arbitrarily misaligned.
}

The canonical unified model for Seyfert galaxies %\cite{anton93}
invokes chance line-of-sight obscuration by a single parsec-scale
torus to explain different observed phenomena.
However, some data are better explained by a model
incorporating a second absorbing
structure\cite{mcleod95,schmitt01,matt00}.
We test this dual-absorber (DA) model,
assuming one absorber is the canonical parsec-scales torus,
arbitrarily misaligned with the host galactic plane,
while the other is at 100-pc scales, aligned with the host galaxy disc
(hereafter the galactic-aligned absorber, or GA).
% A type~2 Seyfert is observed if either absorber lies along the line of
% sight to the central engine.
%
Furthermore, we assume the torus is Compton thick ($N_H >
10^{23.5}$~cm$^{-2}$) and the GA has a lower attenuating column.
Either absorber is capable of obscuring the central engine.

The relative alignment of the two obscuring structures plays an
important role in the DA model.
% , as illustrated by Fig.~\ref{fig:orient}.
% \begin{figure}[ht]
% \centering
% \epsfxsize=7cm   %width of figure - will enlarge/reduce the figures
% \epsfbox{dualabs_orient_fill+key.eps}
% \caption{Illustration of the role played by internal alignment in
%   dual-absorber model.  When the torus and GA are well aligned, most
%   lines of sight 
%   are either obscured by both or neither; when the relative alignment
%   is poor, fewer lines of sight are unattenuated and more see
%   absorption by only the Compton-thin GA.\label{fig:orient}}
% \end{figure}
Compton-thin Seyfert~2s will be observed only if the line of sight
intercepts the GA and avoids the torus.  
If the torus and GA are well aligned much or all of the GA will lie
within the shadow cast by the torus, so randomly-oriented lines of
sight are likely to either intercept both or neither, hence most
well-aligned Seyferts would be observed as either unabsorbed type~1s
or Compton-thick type~2s.
On the other hand, when the absorbers are misaligned the GA covers
part of the opening of the torus, leaving fewer sight lines with a
direct view of the nucleus and more sight lines with Compton-thin
(GA-only) absorption.

\begin{figure}[t]
\centering
% \epsfysize=6.4cm   %width of figure - will enlarge/reduce the figures
% \rotatebox{270}{\epsfbox{feEW_vs_delta-BW.ps}}
\rotatebox{270}{\resizebox{!}{6.4cm}{\includegraphics{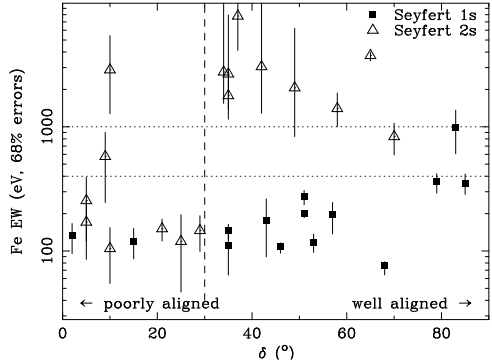}}}
\caption{Compton-thick sources have EW
  $>$ 1~keV, and Compton-thin ones have EW $<$ 400~eV.
  As predicted, Compton-thin Seyfert~2s (GA-only absorption) are only
  found in strongly misaligned systems.
  The distribution of all Compton-thin sources (incl.\ type~1s) is
  insensitive to $\delta$, as expected if only
  the orientation of the torus is important.  
%  Data from \textsl{ASCA} archive.
  \label{fig:EWdelta}}
\end{figure}

To test the DA model, we choose Seyferts for which the
relative alignment can be constrained and use X-ray spectra to
discriminate between Compton-thick and Compton-thin attenuation.
Alignment is indicated by the $\delta$ values\footnote{$\delta$ is
the angle between the (projected) radio jet
and major axis of the host galaxy.}
of Kinney {\it et al.}\cite{kinney00};
small % $\delta$
values indicate strong misalignments.
Fe~K$\alpha$ line equivalent width is measured to constrain absorbing
columns\footnote{Fe~K$\alpha$ EW provides a more
  robust indicator than
  measuring $N_H$ from continuum modeling because values
  outside the range $10^{22}$--$10^{24}$~cm$^{-2}$ are
  not well constrained.}.
Our measurements (Fig.~\ref{fig:EWdelta}) 
%    A KS test shows the likelihood that Seyfert~1s and Compton-thin
%    Seyfert~2s have distinct $\delta$ distributions is $> 99.8\%$.
match the predictions of the DA model.

\end{document}